\documentclass[letterpaper, 10 pt, conference]{ieeeconf}  

\IEEEoverridecommandlockouts                              

\usepackage{graphicx} 
\usepackage{mathptmx} 
\usepackage{amsmath} 
\usepackage{amssymb}  
\usepackage{tikz}
\usetikzlibrary{positioning, fit, backgrounds, arrows.meta, calc}
\usepackage{booktabs}

\title{\LARGE \bf
Robust Scenario-Based Data-Enabled Predictive Control of a Battery Energy Storage System: An Experimental Study*
}

\author{Sebastian Zieglmeier$^{1,2}$, Nikolas Recke$^{1}$ and Mathias Hudoba de Badyn$^{1}$
\thanks{*S.~Zieglmeier was supported by UiO:Energy and Environment at the University of Oslo and acknowledges Empa, the Swiss Federal Laboratories for Materials Science and Technology, for providing the research environment used in this work. N.~Recke and M.~Hudoba de Badyn are supported by FME Solar, funded by the Research Council of Norway under project number 350244.}
\thanks{$^{1}$Sebastian Zieglmeier, Nikolas Recke and Mathias Hudoba de Badyn are with the Department of Technology Systems, University of Oslo, 2027 Kjeller, Norway
        {\tt\small sebastiz@uio.no, nikolalr@uio.no, mathihud@uio.no}}%
\thanks{$^{2}$Sebastian Zieglmeier is with the Urban Energy Systems Laboratory, Swiss Federal Laboratories for Materials Science and Technology (Empa), 8600 Duebendorf, Switzerland}%
}%

\begin{document}

\maketitle
\thispagestyle{empty}
\pagestyle{empty}

\begin{abstract}
Battery energy storage systems must respect strict state-of-charge (SOC) limits to prevent overcharge and deep discharge, a task complicated by measurement noise and costly-to-model nonlinear dynamics. 
Data-enabled predictive control (DeePC) resolves the costly modeling issue by predicting future behavior directly from data. However, its regularization robustifies only the prediction and not the constraints.
Scenario-based DeePC (Scenario-DeePC) extends DeePC with the scenario approach, building constraint robustness fully data-driven from observed prediction errors rather than an assumed disturbance distribution. 
This paper presents the first real-world deployment of Scenario-DeePC, on a grid-connected battery system at the NEST research facility, whose SOC estimate exhibits abrupt, irregular recalibration jumps in addition to ordinary noise. 
Compared to standard DeePC, Scenario-DeePC achieves comparable tracking performance with substantially fewer constraint violations. Its adaptive scenario buffer further tightens constraint handling automatically, improving robustness to unpredictable recalibration events.
\end{abstract}

\section{INTRODUCTION}
Battery energy storage systems are a central enabling technology for the ongoing transition toward renewable and decentralized electricity grids. By decoupling the timing of electricity generation from consumption, grid-connected batteries provide the flexibility required to absorb the intermittency of renewable generation. They further enable load shifting and peak shaving, and support grid balancing services such as frequency regulation. 
As batteries are increasingly deployed not as isolated units but as flexible assets embedded within building- and district-level electric grids, their control must reconcile economic or operational objectives with strict physical operating limits~\cite{recke2026distributionally}.

The relationship between applied power, SOC, and terminal voltage is nonlinear, and cell behavior varies with operating point and state of health. Capturing these effects in an explicit parametric model is in general costly in practice: first-principles modeling requires an accurate characterization of these nonlinear and operating-point-dependent dynamics, while system identification demands extensive, carefully designed experiments to obtain a precise model~\cite{zieglmeier2025semi}. 
Both routes are therefore time-intensive and often impractical to repeat across cells, applications, or aging states. 
Data-enabled predictive control (DeePC)~\cite{coulson2019data} sidesteps this difficulty entirely by predicting future behavior directly from measured input-output trajectories, making it a natural candidate for battery control. No explicit battery model is required, while the reference trajectory can still be tracked in an optimal manner with constraints directly included in the control formulation.

Standard DeePC is built on Willems' fundamental lemma and, for linear time-invariant (LTI) systems, admits established theoretical guarantees on prediction accuracy and closed-loop stability~\cite{berberich2020data, berberich2025overview}. Extending this behavioral framework to nonlinear systems has been pursued along two lines: by generalizing the fundamental lemma itself to linear parameter-varying (LPV) representations~\cite{verhoek2021data, verhoek2025behavioural}, or by modifying the data-driven predictor within the DeePC problem to directly accommodate nonlinear dynamics~\cite{zieglmeier2025gain, naf2025choose, beerwerth2025morecontextualsamplingnonlinear, engeln2026data}. For weaker nonlinearities, regularized DeePC instead absorbs the resulting model mismatch implicitly within the prediction step, an approach that has proven effective in practice~\cite{elokda2021data, zieglmeier2025auv}.

However, the regularization terms in standard DeePC render the data-driven prediction step robust to noise and mild nonlinearities, but do not provide the optimization problem's constraints with any explicit robustness guarantee against uncertainty. 
Robustified variants of DeePC that directly target this gap have since been proposed, including distributionally robust DeePC~\cite{coulson2021distributionally} and scenario-based DeePC (Scenario-DeePC)~\cite{zieglmeier2026scenario}, the latter of which we build on within this paper. Scenario-DeePC extends DeePC with the scenario approach, providing a straightforward, sampling-based way of handling chance constraints within the optimization problem. In a real-world battery system, the SOC must remain within its admissible bounds at all times to prevent overcharge and deep discharge, despite the presence of measurement noise, further real-world disturbances, and residual nonlinear dynamics~\cite{behrunani2023degradation}. Scenario-DeePC's ability to handle such effects directly from data, without requiring an assumed noise or disturbance distribution, while still enabling a fully data-driven approach, makes it the natural choice for the battery control problem considered in this paper.

The contribution of this paper is an experimental validation of Scenario-DeePC on a real-world battery energy storage system connected to the electric grid of the Next Evolution in Sustainable Building Technologies (NEST) research facility in Switzerland, tracking a cyclic SOC reference trajectory. We compare the closed-loop performance of Scenario-DeePC against standard DeePC on this system, quantifying the resulting improvement in constraint satisfaction under real operating conditions.

This paper is organized as follows. 
Section~\ref{sec:method} introduces the Scenario-DeePC foundations. 
Section~\ref{sec:application} presents the battery system, Section~\ref{sec:results} discusses the experimental results, and Section \ref{sec:conc} provides conclusions and directions for future work.

\section{SCENARIO-DEEPC}\label{sec:method}
Consider a discrete-time LTI system with input $u_k \in \mathbb{R}^{n_u}$ and output $y_k \in \mathbb{R}^{n_y}$. A dynamical system is formally defined as $\Sigma = (\mathbb{T}, \mathbb{W}, \mathcal{B})$, where $\mathbb{T}$ is the time axis, $\mathbb{W}$ the signal space, and $\mathcal{B} \subseteq \mathbb{W}^\mathbb{T}$ the behavior of the system. Willems' fundamental lemma~\cite{willems2005note} states that, for a controllable LTI system, a sufficiently long and persistently exciting input-output trajectory $(u^d, y^d)$ spans the entire behavior of the system. A previously measured trajectory of length $T_d$ is therefore arranged into Hankel matrices $\mathcal{H}_L(u^d)$ and $\mathcal{H}_L(y^d)$ of depth $L = T_\text{ini} + N$, partitioned into past and future blocks
\begin{equation}
    \begin{bmatrix} U_p \\ U_f \end{bmatrix} = \mathcal{H}_L(u^d), \quad
    \begin{bmatrix} Y_p \\ Y_f \end{bmatrix} = \mathcal{H}_L(y^d),
\end{equation}
where $U_p, Y_p$ cover the past horizon $T_\text{ini}$ and $U_f, Y_f$ the future horizon $N$. Given the most recent input-output measurements $u_\text{ini}, y_\text{ini}$, DeePC~\cite{coulson2019data} predicts and optimizes the future trajectory directly from data, without an intermediate parametric model:
\begin{subequations}\label{eq:DeePC}
\begin{align}
\min_{g,\,u,\, y,\,\sigma_y}\quad
& \|r - y\|_Q^2 + \|u\|_R^2 + \|g\|_{\lambda_g}^2 + \|\sigma_y\|_{\lambda_y}^2
\label{eq:DeePC_cost}\\
\text{s.t.}\quad
& \begin{bmatrix} U_p \\ Y_p \\ U_f \\ Y_f \end{bmatrix} g
= \begin{bmatrix} u_\text{ini} \\ y_\text{ini} \\ u \\ y \end{bmatrix}
+ \begin{bmatrix} 0 \\ \sigma_y \\ 0 \\ 0 \end{bmatrix}
\label{eq:deepc_eq}\\
& u_k \in \mathcal{U}, \; y_k \in \mathcal{Y}, \quad \forall\, k \in \{0,\ldots,N-1\},
\label{eq:deepc_constr}
\end{align}
\end{subequations}
where $g$ parameterizes admissible trajectories as a linear combination of Hankel columns, $r$ is the reference trajectory, and $Q \succ 0$, $R \succ 0$ weight tracking error and control effort, respectively. The regularization terms $\|g\|_{\lambda_g}^2$ and $\|\sigma_y\|_{\lambda_y}^2$ relax the exact trajectory-matching condition of the fundamental lemma and improve robustness of the prediction with respect to noisy data. They do not, however, provide any explicit robustness guarantees for the constraints $u_k \in \mathcal{U}$, $y_k \in \mathcal{Y}$ against uncertainty in this prediction.

The scenario approach~\cite{calafiore2006scenario, campi2008exact} provides such probabilistic guarantees by offering tractable reformulations of chance-constrained optimization problems. Consider a generic program
\begin{equation}
    \min_{x \in \mathcal{X}} \; c(x) \quad \text{s.t.} \quad \mathbb{P}\{f(x, \delta) \leq 0\} \geq 1 - \varepsilon,
\end{equation}
where $x \in \mathcal{X} \subseteq \mathbb{R}^{n_\text{opt}}$ is the decision variable, with $n_\text{opt}$ the number of decision variables, $c(x)$ a convex cost, $\delta$ an uncertain parameter drawn from a distribution $\mathbb{P}$, and $\varepsilon \in (0,1)$ a prescribed violation tolerance. This chance-constrained program is generally intractable, since the feasible set is not convex for an arbitrary $\mathbb{P}$. The scenario approach replaces the probabilistic constraint with $N_\text{scen}$ independently sampled constraint instances $\{\delta^{(i)}\}_{i=1}^{N_\text{scen}} \sim \mathbb{P}$,
\begin{equation}
    \min_{x \in \mathcal{X}} \; c(x) \quad \text{s.t.} \quad f(x, \delta^{(i)}) \leq 0, \; i = 1, \dots, N_\text{scen},
\end{equation}
yielding a standard convex program. For a confidence parameter $\beta \in (0,1)$, the optimal solution $x^\star_{N_\text{scen}}$ satisfies the probabilistic guarantee $\mathbb{P}^{N_\text{scen}}\{V(x^\star_{N_\text{scen}}) > \varepsilon\} \leq \beta$ whenever
\begin{equation}
    N_\text{scen} \geq \frac{2}{\varepsilon}\left(\ln\frac{1}{\beta} + n_\text{opt}\right), \label{eq:scenario_bound}
\end{equation}
where $V(x^\star_{N_\text{scen}})$ is the violation probability of the obtained solution. This bound is distribution-free: it holds for any $\mathbb{P}$ and requires no structural assumption on the uncertainty, such as Gaussianity or boundedness. The scenario approach has further been reviewed and extended in~\cite{campi2009scenario, campi2021scenario}.

Scenario-DeePC~\cite{zieglmeier2026scenario} extends DeePC with the scenario approach by treating the output constraint in~\eqref{eq:deepc_constr} as a chance constraint and enforcing it over a set of sampled scenarios. Each scenario perturbs the predicted output $y_\text{pred} = Y_f g$ by a sampled realization $s^{(i)} \in \mathbb{R}^{n_y N}$,
\begin{equation}
    y^{(i)}_k = y_{\text{pred},k} + s^{(i)}_k, \quad k = 0, \dots, N-1,
\end{equation}
and the constraint $y_k^{(i)} \in \mathcal{Y}$ is enforced for every scenario $i = 1, \dots, N_\text{scen}$, while the scenario outputs are likewise incorporated into the tracking cost. Rather than assuming a distribution for the disturbance $\delta$, the scenarios are constructed directly from data, in line with the data-driven philosophy underlying DeePC: at each closed-loop step, the one-step prediction error
\begin{equation}
    w = y - y_{\text{pred},0}
\end{equation}
between the measured output $y$ and the predicted output $y_{\text{pred},0}$ is computed and stored in a buffer $\mathcal{E}_\text{buffer}$. This buffer thus represents the empirical distribution of the uncertainty actually acting on the closed-loop system, rather than a distribution assumed a priori. Scenarios are then drawn by sampling, independently for each horizon step, from this empirical buffer, $s_k^{(i)} \sim \text{Unif}(\mathcal{E}_\text{buffer})$.

Since consecutive prediction errors are temporally correlated through the closed-loop dynamics, while the scenario guarantee of~\eqref{eq:scenario_bound} requires independent samples, only every $(M+1)$-th prediction error is retained in the buffer, where $M$ is the correlation horizon of the underlying $M$-dependent error process. This decorrelation step recovers a buffer of independent, identically distributed samples of the empirical error distribution, so that the classical scenario bound~\eqref{eq:scenario_bound} applies directly to the resulting Scenario-DeePC problem, yielding a distribution-free probabilistic guarantee on output constraint satisfaction~\cite{zieglmeier2026scenario}.

Since tightened scenario constraints can render~\eqref{eq:deepc_constr} infeasible, a non-negative slack variable $h_k^{(i)} \in \mathbb{R}^{n_y}_{\geq 0}$ is introduced for each scenario and time step to relax the output constraint set component-wise, restoring feasibility at every time step~\cite{zieglmeier2026scenario}. Each slack is penalized quadratically in the cost with weight $\mu > 0$. This results in the following final formulation
\begin{subequations}\label{eq:ScDeePC_relaxed}
\begin{align}
\min_{g,\,u,\, y_\text{pred},\,\sigma_y,\,\{h_k^{(i)}\}}\quad
& \begin{aligned}[t]
    & \frac{1}{N_\text{scen}} \sum_{i=1}^{N_\text{scen}} \Big(\|r - y^{(i)}\|_Q^2 + \|h^{(i)}\|_ \mu^2\Big) \\
    & \quad + \|u\|_R^2 + \|g\|_{\lambda_g}^2 + \|\sigma_y\|_{\lambda_y}^2
  \end{aligned}
\label{eq:ScDeePC_relaxed_cost}\\
\text{s.t.}\quad
& \begin{bmatrix} U_p \\ Y_p \\ U_f \\ Y_f \end{bmatrix} g
= \begin{bmatrix} u_\text{ini} \\ y_\text{ini} \\ u \\ y_\text{pred} \end{bmatrix}
+ \begin{bmatrix} 0 \\ \sigma_y \\ 0 \\ 0 \end{bmatrix}
\label{eq:ScDeePC_relaxed_data}\\
& y_k^{(i)} = y_{\text{pred},k} + s_k^{(i)}
\label{eq:ScDeePC_relaxed_scenario}\\
& h_k^{(i)} \geq 0 \\
& u_k \in \mathcal{U} \quad y_k^{(i)} \in \mathcal{Y}(h_k^{(i)})
\label{eq:ScDeePC_relaxed_constraints}\\
& \forall\, k \in \{0,\ldots,N-1\}\;
\forall\, i \in \{1,\ldots,N_\text{scen}\}, \notag
\end{align}
\end{subequations}
where $h^{(i)} := (h_0^{(i)}, \ldots, h_{N-1}^{(i)})$ stacks the per-step slacks over the horizon, $\underline{y}$ and $\overline{y}$ denote the lower and upper output constraint bounds from $\mathcal{Y}$, and $\mathcal{Y}(h^{(i)}) := \{y \in \mathbb{R}^{n_y} \mid \underline{y} - h^{(i)} \leq y \leq \overline{y} + h^{(i)}\}$ denotes the output constraint set relaxed by $h^{(i)}$.

In this paper, the adaptive variant of Scenario-DeePC is employed: rather than being populated offline prior to control operation, the buffer $\mathcal{E}_\text{buffer}$ is updated online during closed-loop operation. At every $(M+1)$-th closed-loop step, the newly computed prediction error is added to the buffer while the oldest stored error is discarded, so that $\mathcal{E}_\text{buffer}$ continuously reflects the most recent operating conditions of the system. This allows the constraint tightening imposed by the scenario formulation to adapt automatically to changes in measurement noise, disturbances, or operating-point-dependent model mismatch encountered during operation.

\section{APPLICATION}\label{sec:application}
The considered battery energy storage system (BESS) is an ads-tec storage rack system, installed at NEST, a research and innovation building operated by Empa in Dübendorf, Switzerland. The BESS is connected to NEST's electric grid through a bidirectional inverter. A storage rack controller aggregates ten storage rack battery modules, each built from LeClanché lithium titanate cells, into a single virtual battery. This aggregated battery is exposed to the controller over a Modbus/TCP interface, so the underlying multi-module structure is not directly accessible. As with any electrochemical energy storage system, the BESS's usable capacity and internal characteristics degrade with cycling and calendar aging, such that its present state of health differs from its condition at commissioning.

\begin{figure}
    \centering
    \includegraphics[width=0.7\linewidth]{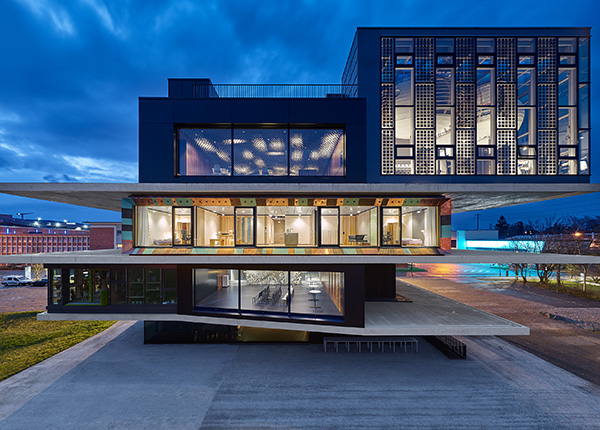}
    \caption{The NEST research facility. \copyright Zooey Braun.}
    \label{fig:NEST}
\end{figure}

The BESS is characterized by its terminal voltage $v_k$, current $i_k$, and the resulting instantaneous power $p_k$, which relate as
\begin{equation}
    p_k = v_k\, i_k.
\end{equation}
The state of charge evolves as a function of the charging and discharging current exchanged with the grid. For a time step of duration $\Delta t$, the underlying relationship between current and SOC is given by
\begin{equation}
    SOC_{k+1} = SOC_k + \frac{\Delta t}{Q_\text{cap}}\left(\eta_\text{ch}\, i_{\text{ch},k} - \eta_\text{dis}^{-1} i_{\text{dis},k}\right),
\end{equation}
where $SOC_k \in [0,1]$ denotes the SOC, $Q_\text{cap}$ the charge capacity of the battery, and $i_{\text{ch},k}, i_{\text{dis},k} \geq 0$ the charging and discharging current, respectively, with $\eta_\text{ch}, \eta_\text{dis} \in (0,1]$ the corresponding conversion efficiencies. Charging and discharging are mutually exclusive at any time step.

The energy stored in the battery is $E_k = SOC_k\, E_\text{cap}$, where $E_\text{cap}$ denotes the nominal energy capacity. This stored energy evolves as a function of the charging and discharging power exchanged with the grid,
\begin{equation}
    E_{k+1} = E_k + \Delta t\left(\eta_\text{ch}\, p_{\text{ch},k} - \eta_\text{dis}^{-1} p_{\text{dis},k}\right),
\end{equation}
so that, combined with the relationship above,
\begin{equation}
    SOC_{k+1} = SOC_k + \frac{\Delta t}{E_\text{cap}}\left(\eta_\text{ch}\, p_{\text{ch},k} - \eta_\text{dis}^{-1} p_{\text{dis},k}\right),
\end{equation}
directly relating the charging and discharging power to the evolution of the SOC.

The BESS has a nominal energy capacity of 96 kWh, a rated voltage range of 600 to 840 V, and a maximum charge and discharge current of 300 A. To protect the cells, the system is operated within a restricted SOC window of 20\% to 80\%, rather than over its full nominal range.

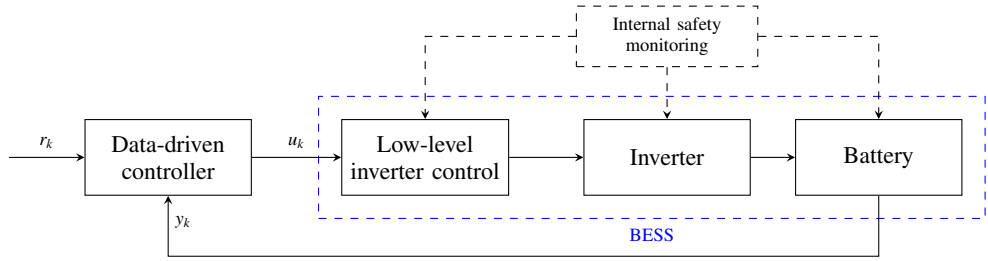
\begin{figure*}[t]
    \centering
    \begin{tikzpicture}[
        block/.style={draw, rectangle, minimum height=1cm, minimum width=2.2cm, align=center, font=\small},
        sblock/.style={draw, rectangle, minimum height=0.8cm, minimum width=2.4cm, align=center, font=\scriptsize, dashed},
        >=stealth
    ]
        \node[block] (ctrl) at (0,0) {Data-driven\\controller};
        \node[block] (llc)  at (3.4,0) {Low-level\\inverter control};
        \node[block] (inv)  at (6.6,0) {Inverter};
        \node[block] (bat)  at (9.4,0) {Battery};

        \node[draw, dashed, blue, fit={(llc) (inv) (bat)}, inner sep=0.3cm] (bess) {};
        \node[font=\scriptsize, blue] at (bess.south) [below] {BESS};

        \node[sblock] (safety) at (6.6,1.6) {Internal safety\\monitoring};

        \draw[->] ($(ctrl.west)+(-1.0,0)$) -- node[above, font=\scriptsize]{$r_k$} (ctrl.west);

        \draw[->] (ctrl) -- node[above, font=\scriptsize]{$u_k$} (llc);
        \draw[->] (llc) -- (inv);
        \draw[->] (inv) -- (bat);

        \draw[->] (bat.south) -- ++(0,-0.8) -| node[below, pos=0.97, font=\scriptsize, yshift=-1mm, xshift=2mm]{$y_k$} (ctrl.south);

        \draw[dashed, ->] (safety.west) -| (llc.north);
        \draw[dashed, ->] (safety.south) -- (inv.north);
        \draw[dashed, ->] (safety.east) -| (bat.north);
    \end{tikzpicture}
    \caption{Control architecture: a data-driven controller commands a power setpoint $u_k$ to the BESS, tracking a reference $r_k$ from the measured SOC $y_k$. The low-level inverter control, inverter, and battery are treated jointly as a single black-box system (BESS). An internal safety mechanism monitors and, if necessary, disconnects the battery independently of the external controller.}
    \label{fig:architecture}
\end{figure*}

Charge and discharge power is commanded and measured at the output of the inverter rather than directly at the battery terminals. The manipulated input $u_k$ therefore corresponds to the commanded inverter power at the point of grid connection, while the measured output $y_k$ corresponds to the estimated SOC of the aggregated battery. Since $u_k$ and $y_k$ are defined at this interface rather than at the cells themselves, electrical losses, estimation effects, and dynamics introduced by the inverter and its connection to the battery are inherently reflected in the measured data.

The SOC of a battery is not directly observable, but is instead estimated by the battery management system from current and voltage measurements~\cite{brandle2026optimal}, a characteristic common to BESSs in general. In this work, the underlying estimation algorithm of the BESS is treated as a black box. The battery consists of multiple modules, each in turn composed of multiple cells, each of which is individually estimated internally. The system aggregates these individual estimates into the single SOC value made available to the controller, without disclosing how this aggregation is performed. The controller in this work accesses the system at the aggregated system level, and therefore observes only a single estimated system SOC value, without visibility into the underlying per-module or per-cell estimates.

In closed-loop operation, this estimated SOC is affected by two distinct sources of uncertainty. First, it exhibits continuous variation consistent with ordinary estimation noise and limited measurement resolution. Second, the estimate occasionally undergoes an internal recalibration, causing an abrupt, discrete jump of several percentage points, in either direction, at irregular and a priori unknown points in time, distinct from the SOC's ordinary evolution under charging and discharging.

Operating a battery outside its safe voltage range, either by overcharging or by discharging it too deeply, accelerates cell degradation and, in extreme cases, can cause irreversible damage or safety hazards. To prevent this, the BESS continuously monitors cell voltages and temperatures internally. If a critical threshold is exceeded, the system autonomously opens its internal DC contactors, disconnecting the battery from the inverter and reporting an error state, irrespective of any command issued by the external controller.

\section{RESULTS}\label{sec:results}
This section presents the experimental validation of Scenario-DeePC on the battery system described in Section~\ref{sec:application}, covering the experimental setup, the hyperparameter tuning procedure, the resulting choice of correlation horizon $M$, and the numerical comparison against standard DeePC.
\begin{figure*}[t]
    \centering
    \includegraphics[width=1\linewidth]{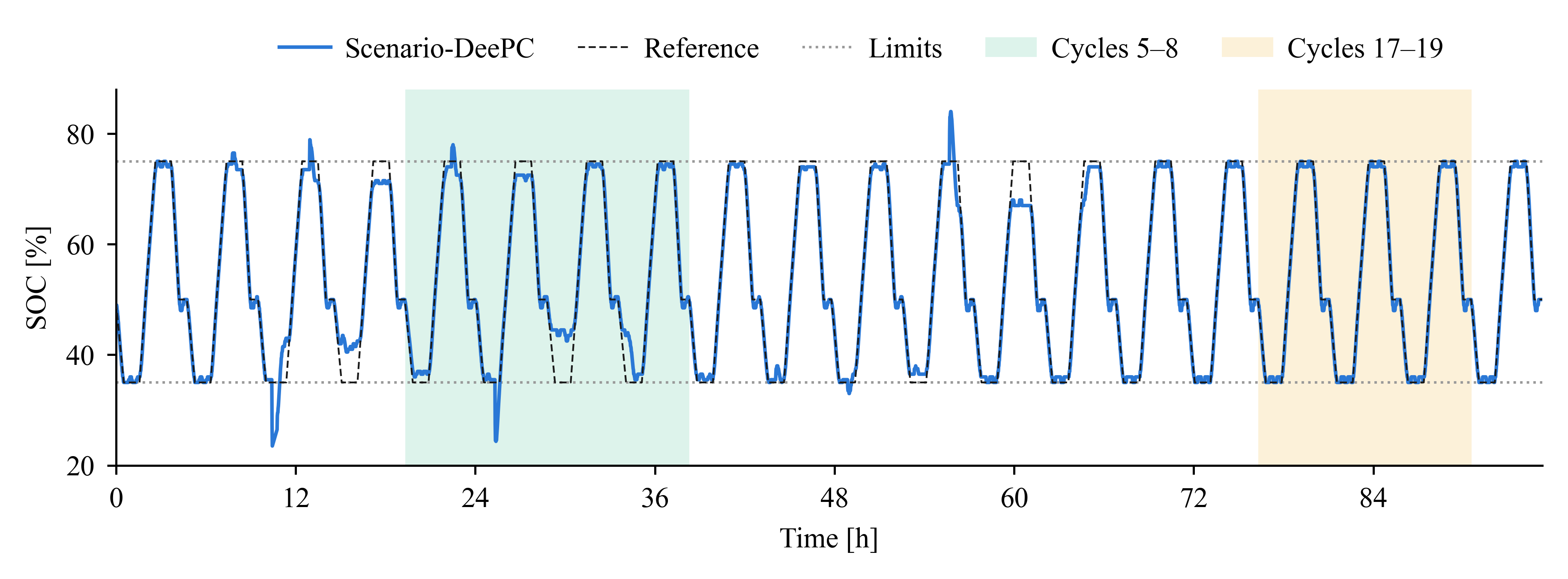}
    \caption{SOC of Scenario-DeePC over all 20 cycles of the reference trajectory. Recalibration events in the estimated SOC are visible as abrupt, irregular jumps. The highlighted windows indicate the cycles shown in detail in Figs.~\ref{fig:standard} and~\ref{fig:calibration}.}
    \label{fig:overview}
\end{figure*}
\subsection{Experimental Setup}\label{sec:setup}
Fig.~\ref{fig:architecture} shows the control architecture used in the experiment. At each control interval, the data-driven controller computes a power setpoint $u_k$ from the estimated SOC and commands it to the BESS. A low-level inverter control loop, internal to the system and operating independently of the external controller, tracks this setpoint by regulating the inverter accordingly. The inverter and the battery are treated jointly as a single black-box system, in line with the data-driven philosophy of DeePC: only the commanded power setpoint $u_k = p_k$ and the estimated $y_k = SOC_k$ are available to the controller, with no access to the internal dynamics of either component outlined in Section~\ref{sec:application}. 
Since the NEST research facility logs and provides data at a resolution of one minute, the controller is operated at a sampling time of one minute as well, solving the underlying optimization problem once per interval. Before the control loop starts, the past-data buffer required by DeePC is prefilled from historical system data, so the controller is operational from the first control step rather than requiring an initial warm-up phase.

The SOC values used throughout this work refer to the true, unscaled state of charge of the battery cells, as reported directly by the research unit, rather than to a customer-facing scale. This differs from the SOC displayed in most commercial battery systems, including electric vehicles, where the displayed 0\% to 100\% range is intentionally mapped onto a safety-buffered range of the cells' true physical range. As a result, a commercial product's displayed 20\% SOC corresponds to a comfortably safe operating point, whereas the same value expressed as a true, unscaled SOC can be dangerously close to deep discharge. On the research unit used in this work, the internal safety monitoring described in Section~\ref{sec:application} disconnects the battery once the true SOC falls below 10\% or rises above 90\%, reflecting the limits beyond which the cells would be damaged. Reflecting this, the reference trajectory tracked by both controllers is a repeating charge-discharge cycle within a further restricted 35\% to 75\% SOC window, chosen to provide an additional margin against the risk that a sudden SOC recalibration, as described in Section~\ref{sec:application}, would otherwise trigger this safety disconnect. The reference trajectory consists of a hold at 50\% SOC, a ramp down to 35\%, a hold, a ramp up to 75\%, a further hold, and a ramp back down to 50\%. This cycle is repeated multiple times over the course of the experiment, simulating the use of the battery across its full operating range.

Both controllers enforce the input constraint set $\mathcal{U} = \{u \in \mathbb{R} \mid \underline{u} \leq u \leq \overline{u}\}$ and the output constraint set $\mathcal{Y} = \{y \in \mathbb{R} \mid \underline{y} \leq y \leq \overline{y}\}$, with $\underline{u} = -50\,\text{kW}$, $\overline{u} = 50\,\text{kW}$, and $\underline{y} = 35\%$, $\overline{y} = 75\%$, matching the SOC window of the reference trajectory. The reference trajectory is deliberately chosen to reach these constraint bounds rather than remain strictly within them, since only in this regime does uncertainty in the predicted output risk an actual constraint violation, motivating the chance-constraint scenario-based formulation in the first place~\cite{zieglmeier2026scenario}.

The data-driven controllers were implemented in \texttt{Python 3.14.3}, with the underlying optimal control problems formulated in \texttt{CVXPY 1.8.2} and solved using \texttt{MOSEK 11.1.10}. Communication with the BESS was established via the MQTT protocol.

The data underlying the Hankel matrices was collected from the system's normal closed-loop operation on 26 March 2026, 10:00 to 17:58, yielding $T_d = 479$ ($460$ Hankel columns). This window was chosen close in time to the experiment itself, conducted from 21 to 27 April 2026, so that the state of health reflected in the data aligns closely with that of the battery during the experiment.

\subsection{Hyperparameter Tuning}\label{sec:hp-tuning}
\begin{table}[]
    \centering
    \caption{Hyperparameters used for DeePC and Scenario-DeePC.}
    \label{tab:hp}
    \resizebox{\columnwidth}{!}{%
    \setlength{\tabcolsep}{3pt}%
    \begin{tabular}{lcccccccccc}
        \toprule
         & $T_\text{ini}$ & $N$ & $Q$ & $R$ & $\lambda_g$ & $\lambda_y$ & $\mu$ & $N_\text{scen}$ & $N_\text{buffer}$ \\
        \midrule
        DeePC & 10 & 10 & $10^3$ & $10^{-2}$ & $10^6$ & $10^5$ & -- & -- & -- \\
        Scenario-DeePC & 10 & 10 & $10^3$ & $10^{-2}$ & $10^6$ & $10^5$ & $10^6$ & 100 & 500 \\
        \bottomrule
    \end{tabular}%
    }
\end{table}

Tuning hyperparameters directly on the physical system through a full search is impractical. Each control step commits a real charge or discharge action on the battery, and the effect of a hyperparameter change only becomes visible over several control intervals. The degrees of freedom were therefore reduced before deployment, following the same procedure as~\cite{zieglmeier2026modelfree}: the data-driven predictor was first calibrated offline as a pure prediction model, using recorded charge and discharge power and the corresponding measured SOC from prior operation of the system. Given this historical input, the SOC was predicted over the prediction horizon without solving for an optimal input, and the regularization weights $\lambda_g$ and $\lambda_{y}$ were tuned to maximize prediction accuracy. This fixed the two weights relative to one another, leaving only their ratio to $Q$ and $R$ to be tuned during deployment. The scenario parameters $N_\text{scen}$ and $N_\text{buffer}$ were fixed a priori, while the scenario buffer itself was updated online during operation, as described in Section~\ref{sec:method}. The resulting hyperparameters are listed in Table~\ref{tab:hp}.

\subsection{M-Dependence in Practice}\label{sec:M}
Assumption~3 in~\cite{zieglmeier2026scenario} requires the buffer $\mathcal{E}_\text{buffer}$ to consist of effectively independent prediction errors, achieved by retaining only every $(M+1)$-th error. In this deployment, however, the one-step prediction errors $w = y - y_{\text{pred},0}$ are highly correlated. We attribute this primarily to the limited resolution of the estimated SOC, which is provided by the battery management system in steps of 0.5\%, consistent with prior work on this battery system~\cite{yin2024data}.
This quantization affects both terms of the prediction error: the measured SOC $y$ is itself quantized, and the data-driven predictor, built exclusively from measured trajectories of the same resolution, cannot resolve SOC variations below 0.5\% either. During the hold segments of the reference trajectory, the SOC changes by considerably less than 0.5\% per control interval. Therefore, the resulting quantization error does not vary independently from step to step but persists over several consecutive samples. 
Enforcing decorrelation through subsampling would therefore require a large $M$. At the one-minute control interval used in this work and a buffer size of $N_\text{buffer} = 500$, admitting only a few new errors into the buffer per hour would require several days to fully populate it and furthermore omit disturbances, relevant for the current operating point.
Since this work focuses on the practical robustification offered by Scenario-DeePC rather than on its theoretical guarantees, we deliberately choose $M = 0$, retaining every prediction error. This lets the buffer fill quickly and allows us to assess experimentally whether the scenario approach provides sufficient robustification even under this correlated, quantization-dominated error process. With this choice, the probabilistic guarantee of Theorem~1 in~\cite{zieglmeier2026scenario} does not formally apply; the resulting robustness is instead evaluated experimentally.

\subsection{Experimental Results}\label{sec:experimental}
\begin{figure}[t]
    \centering
    \includegraphics[width=1\linewidth]{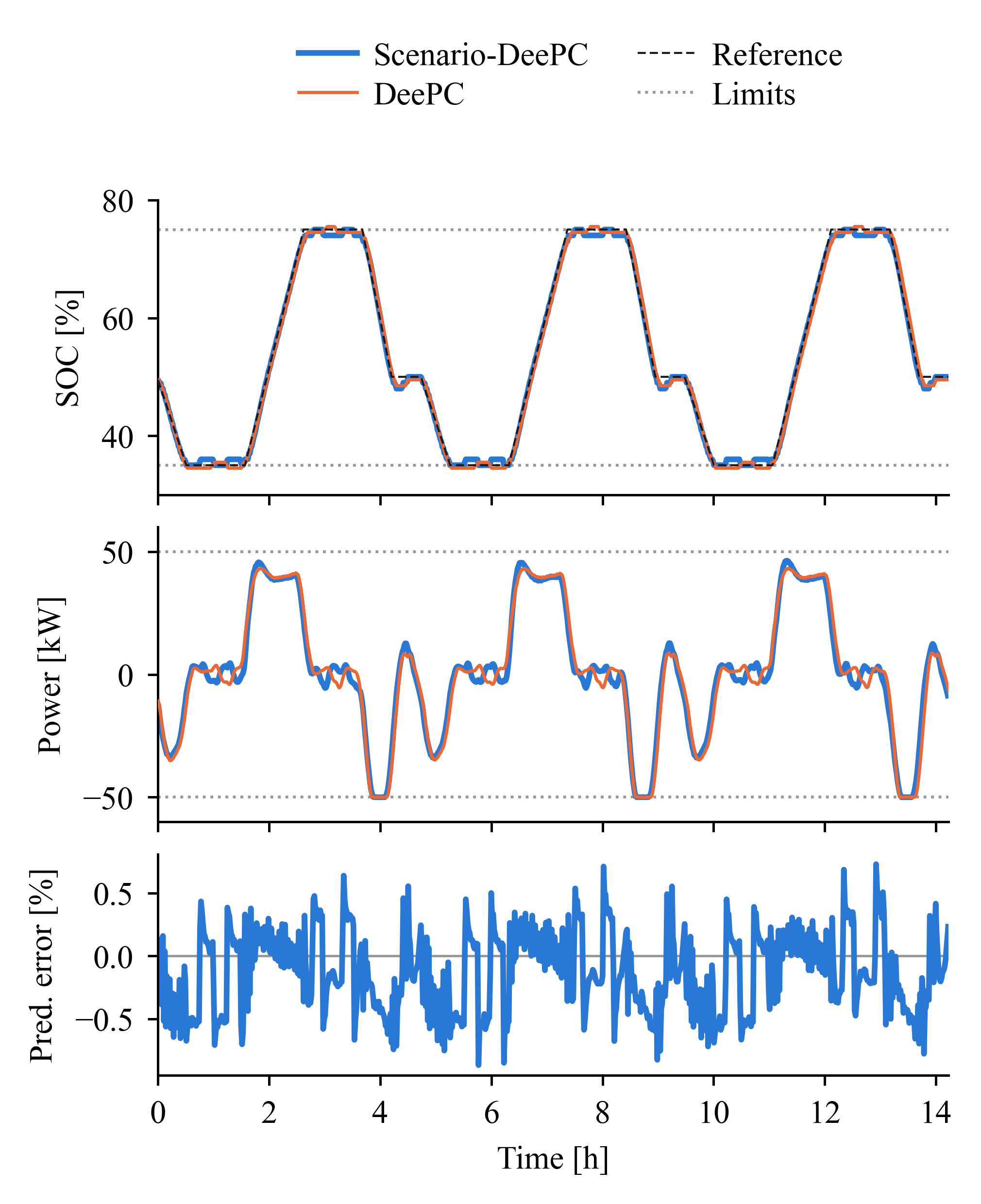}
    \caption{SOC, commanded power, and one-step prediction error of Scenario-DeePC and standard DeePC during cycles of the reference trajectory without a recalibration event, shown superimposed in time.}
    \label{fig:standard}
\end{figure}

\begin{figure}[t]
    \centering
    \includegraphics[width=1\linewidth]{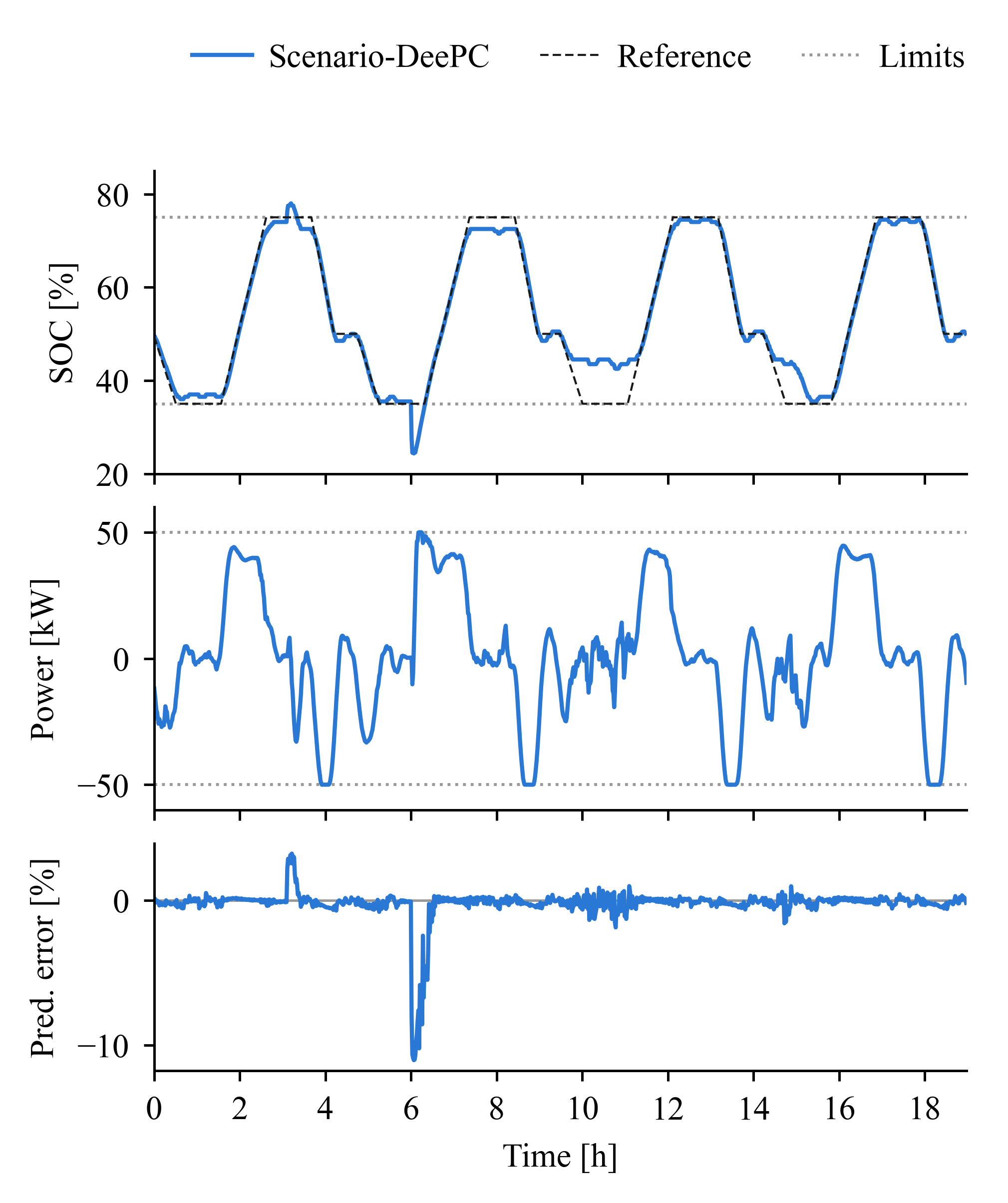}
    \caption{SOC, commanded power, and one-step prediction error of Scenario-DeePC over consecutive cycles of the reference trajectory containing two recalibration events of opposite sign. The elevated scenario buffer entries introduced by these events visibly tighten the constraint handling for the subsequent cycles, until discarded from the buffer.}
    \label{fig:calibration}
\end{figure}

Fig.~\ref{fig:overview} shows the SOC of Scenario-DeePC over the full experiment, spanning all 20 cycles of the reference trajectory. Several SOC recalibration events, as described in Section~\ref{sec:application}, are visible as abrupt jumps in the estimated SOC, occurring at irregular and unpredictable points in time and of varying magnitude and direction. Despite these events, Scenario-DeePC succesfully completed all 20 cycles of the reference trajectory. Under the same conditions, standard DeePC failed to complete the experiment, deviating from the admissible SOC range after 7 cycles. The highlighted windows in Fig.~\ref{fig:overview} indicate the cycles examined in detail in Figs.~\ref{fig:standard} and~\ref{fig:calibration}.

To isolate the effect of the scenario-based constraint handling from the effect of the SOC recalibration events, we first compare Scenario-DeePC and standard DeePC on cycles of the reference trajectory during which no SOC recalibration event occurred. Since both controllers could not be operated simultaneously on the physical system, the compared cycles were recorded at different times and are shown superimposed in time in Fig.~\ref{fig:standard}.

As shown in Fig.~\ref{fig:standard}, the two controllers achieve practically identical tracking performance, with a root mean squared error (RMSE) of 0.78\% for both Scenario-DeePC and standard DeePC over the compared windows of 810 samples each. 
In terms of constraint satisfaction, however, standard DeePC exhibits 158 constraint violations, compared to none for Scenario-DeePC. 
All DeePC violations occur at points where the reference itself lies exactly on a constraint bound, with the majority of violations exceeding it by a single 0.5\% quantization step. 
The one-step prediction error shown in the bottom panel of Fig.~\ref{fig:standard} remains approximately within $\pm 0.5\%$ throughout, matching the SOC measurement resolution. At this resolution, a prediction accuracy beyond $\pm 0.5\%$ is inherently difficult to achieve.
Regardless of whether a given violation stems from this resolution or from other unmodeled system effects, Scenario-DeePC remains robustified against it while standard DeePC does not. This illustrates a central strength of the fully data-driven approach: effects such as measurement resolution are inherently captured within the data used to construct the predictor, and hence within the system dynamics implicitly identified by the controller, without requiring these effects to be modeled explicitly. However, only Scenario-DeePC translates this into actual constraint satisfaction, since the scenario approach robustifies the constraints themselves against the resulting prediction error.

We next examine the behavior of the adaptive Scenario-DeePC formulation when a recalibration event does occur. Fig.~\ref{fig:calibration} shows SOC, commanded power, and one-step prediction error of Scenario-DeePC over consecutive cycles of the reference trajectory containing two recalibration events of opposite sign: a moderate upward jump of approximately $3\%$ and, shortly after, a substantially larger downward jump of approximately $10\%$. Both are clearly distinguishable in the prediction error from the ordinary estimation noise observed under standard operation, which remains close to zero by comparison. Each event is registered by the controller as a one-step prediction error and subsequently enters the scenario buffer $\mathcal{E}_\text{buffer}$. It remains available for sampling for $N_\text{buffer} = 500$ control steps, before being discarded due to the rolling buffer update. This is clearly visible in Fig.~\ref{fig:calibration} at around $t = 15$~h, where the controller's tightened constraint handling relaxes as the corresponding buffer entries are discarded.

We highlight three aspects of this behavior. First, the disturbance realizations entering the buffer as a consequence of a recalibration event do not, in general, satisfy the independence assumption underlying the classical scenario approach, nor the $M$-dependence assumption discussed in Section~\ref{sec:hp-tuning}: a recalibration event is a rare, isolated occurrence rather than a repeated draw from a stationary distribution. This is an inherent characteristic of a real-world deployment rather than a theoretical idealization, and it does not prevent the mechanism from providing practical robustification, as observed in Fig.~\ref{fig:calibration}.

Second, the achieved robustification strikes a balance between excessive conservatism and optimal tracking. 
Once a recalibration event enters the buffer, the scenario-based constraint tightening increases visibly for the following cycles. This comes at the cost of reduced tracking accuracy during this period, as the controller trades off how tracking performance for a wider safety margin near the bound. The tightening relaxes again once the corresponding buffer entry is discarded.
The buffer size $N_\text{buffer}$ and the number of scenarios $N_\text{scen}$ directly govern this behavior. Together, they determine both the likelihood that a given recalibration event is drawn as a scenario at a given control step and the duration for which it remains available to be sampled.
Due to the limited experimental time available on the physical system, these parameters were not further fine-tuned specifically for the observed recalibration behavior. Furthermore, the underlying recalibration mechanism itself remains a black box in this work, as discussed in Section~\ref{sec:application}. 

Third, Fig.~\ref{fig:calibration} illustrates this buffer update behavior directly: each recalibration event is captured as an elevated entry in the scenario buffer, visibly robustifying the controller's behavior on the corresponding bound over the following cycles, until the entry is discarded after $N_\text{buffer}$ control steps. A further recalibration event of comparable or smaller magnitude and of the same sign, occurring within this window, would be inherently covered by the already-tightened constraint handling on that bound.

\section{CONCLUSION}\label{sec:conc}
This paper presented the first real-world deployment of Scenario-DeePC, applied to a battery energy storage system at the NEST research facility.
We showed that the approach remains effective on a real, aged, multi-cell system with an undisclosed SOC estimator, even under real-world uncertainty that departs from its underlying theoretical assumptions.
During standard operation, Scenario-DeePC matched the tracking performance of standard DeePC while incurring substantially fewer constraint violations, aligning fully with~\cite{zieglmeier2026scenario}. 
Under SOC recalibration events, largely unaddressed sources of uncertainty in existing battery control approaches, Scenario-DeePC automatically tightened its constraint handling for a limited number of subsequent control steps. This robustified the system against renewed violations until the corresponding buffer entry was discarded.
From a battery perspective, this directly supports safe operation within the admissible SOC range, reducing the risk of overcharge and deep discharge, and thus of accelerated degradation. This is achieved without requiring an explicit model of the battery, the inverter, or the underlying estimation process.
Future work will focus on characterizing the recalibration process itself, to inform a more targeted choice of buffer size and scenario number. Further research is also required to obtain a more practical treatment of the $M$-dependence or on an alternative way to obtain i.i.d.\ scenarios and thereby restore the guarantees of the scenario approach.

{\footnotesize
\textbf{Declaration of Generative AI and AI Technologies in the Writing Process:} During the preparation of this work, the author(s) used Claude (Anthropic) in order to improve the language and readability of this paper. After using this tool/service, the author(s) reviewed and edited the content as needed and take full responsibility for the content of the publication.
}
\bibliographystyle{IEEEtran}
\bibliography{MyBib2}
\end{document}